\documentclass{article}

\usepackage[utf8]{inputenc}
\usepackage[margin=4cm,a4paper]{geometry}
\usepackage{float}
\usepackage{amsmath}
\usepackage[colorlinks]{hyperref}
\usepackage[inline]{enumitem}
\usepackage{color,soul}
\usepackage[svgnames]{xcolor}
  \definecolor{diffstart}{named}{Grey}
  \definecolor{diffincl}{named}{Green}
  \definecolor{diffrem}{named}{OrangeRed}
\usepackage{listings}
\lstdefinelanguage{diff}{
    basicstyle=\ttfamily\small,
    morecomment=[f][\color{diffstart}]{@@},
    morecomment=[f][\color{diffincl}]{+\ },
    morecomment=[f][\color{diffrem}]{-\ },
  }
\definecolor{light-gray}{gray}{0.90}
\sethlcolor{light-gray}
\title{The CodRep Machine Learning on Source Code Competition}
\author{Zimin Chen \and Martin Monperrus}
\date{KTH Royal Institute of Technology\\zimin@kth.se, martin.monperrus@csc.kth.se }

\begin{document}
\maketitle

\begin{abstract}
CodRep is a machine learning competition on source code data. 
It is carefully designed so that anybody can enter the competition, whether professional researchers, students or independent scholars, without specific knowledge in machine learning or program analysis.  
In particular, it aims at being a common playground on which the machine learning and the software engineering research communities can interact.
The competition has started on April 14th 2018 and has ended on October 14th 2018.
The CodRep data is hosted at \url{https://github.com/KTH/CodRep-competition/}.
\end{abstract}

\section{Introduction}

Competitions are great to foster creativity in a problem domain.
The CodRep competition aims at encouraging scientific and technological progress in the domain of machine learning over source code.

The CodRep competition is carefully designed so that anybody can enter the competition, whether professional researcher, student or independent scholar, without specific knowledge in machine learning or program analysis.  
The CodRep competition can also be seen as a common playground on which the machine learning and the software engineering research communities can interact.

The competition requires to build a system that takes as input a set of pairs (source code line, source code file), and outputs, for each pair, the predicted line number of the source code to be replaced by  the source code line.
The participant are given datasets carefully extracted and curated from open-source projects. For instance, \texttt{Dataset1} is composed of 4394 prediction tasks.

The competition starts on April 14th 2018 and ends on Oct 14th 2018.
After the competition, we envision that the curated data provided for CodRep will be used for further empirical studies or as training data for other tasks. 

The competition is organized by KTH Royal Institute of Technology, Stockholm, Sweden and hosted at \url{https://github.com/KTH/CodRep-competition/}.

To sum up, CodRep provides the community with:
\begin{itemize}
\item A novel problem statement for software evolution analysis: given a line and a file, predict where the line is inserted, as well as a loss function specifically designed for this problem.
\item Five curated datasets of one-liner commits from open-source projects. In total, they are composed of 58069 one-liner commits. To our knowledge, this is the largest ever dataset of one-liner commits.

\end{itemize}

\section{Prediction Task}

The competition consists in writing a program which predicts where to insert a specific line into a source code file. In particular, we consider replacement insertions, where the new line replaces an old line, such as \\

\begin{lstlisting}[language=diff]
public class Test {
    int a = 1;
-   int b = 0.1;
+   double b = 0.1;
}
\end{lstlisting}

\noindent
More specifically, the program takes as input a set of pairs (source code line, source code file), and outputs, for each pair, the predicted line number of the line to be replaced by in the initial source code file. \\

\section{Data Structure and Format}

\subsection{Data format}

The provided data are in \hl{Datasets/.../Tasks/*.txt}. The txt files are meant to be parsed by competing programs. Their format is as follows, each file contains:

\begin{lstlisting}
{Code line to insert}
\newline
{The full program file}
\end{lstlisting}

\noindent
For instance, let's consider this example input file, called \hl{foo.txt}.

\begin{lstlisting}[language=java]
double b = 0.1;

public class test{
    int a = 1;
    int b = 0.1;
}
\end{lstlisting}

\noindent
In this example, \hl{double b = 0.1;} is the code line to be added somewhere in the file in place of another line. \\

\noindent
For such an input, a competing program outputs for instance \hl{foo.txt 3}, meaning replacing line 3 (\hl{int b = 0.1;}) with the new code line \hl{double b = 0.1;}. \\

\subsection{Training}

\noindent
To train the system, the correct answer for all input files is given in folder \hl{Datasets/\\.../Solutions/*.txt}, e.g. the correct answer to \hl{Datasets/Datasets1/Tasks/1.txt} is in \hl{Datasets/Datasets1/Solutions/1.txt}

\subsection{Data provenance}

The data used in the competition is taken from real commits in open-source projects.

\textbf{Project selection criterion:} We consider open-source projects that were studied at least in an academic article. The list of considered articles is shown in Table \ref{tab:provenance}.

\textbf{Commit selection criteria:}  For the selected projects, we have analyzed all commits and extracted all the one line replacement changes. We have further filtered the data based on the following criteria:
\begin{enumerate}
\item Only source code files are kept (Java files)
\item Comment-only changes are discarded (e.g. replacing \hl{// TODO} with \hl{// Fixed})
\item Inserted or removed lines are not empty lines, and are not space-only changes
\item Only one replaced code line in the whole file
\end{enumerate}
Table \ref{tab:stats} gives the main descriptive statistics of the dataset.

\begin{table}
\caption{The provenance of each of the 5 datasets used in the competition.}
\label{tab:provenance}
\centering
\begin{tabular}{|l|l|p{7cm}|}
\hline
Dataset &  Source \\ \hline
Dataset1 & \textcolor{blue}{\cite{Zhong2015AnES}}            \\ \hline
Dataset2 & \textcolor{blue}{\cite{monperrus:hal-00769121}}             \\ \hline
Dataset3 & \textcolor{blue}{\cite{li2016watch}}             \\ \hline
Dataset4  & \textcolor{blue}{\cite{Scholtes2016}}       \\ \hline
Dataset5& \textcolor{blue}{\cite{zhou2012should,tufano2017empirical,hata2012bug} }       \\ \hline
\end{tabular}
\end{table}

\begin{table}
\caption{Main descriptive statistics of the CodRep data}
\label{tab:stats}
\centering
\begin{tabular}{|l|l|l|}
\hline
Directory & \# prediction tasks & Total Lines of code (LOC) \\ \hline
Dataset1/ & 3858                    & 2056900             \\ \hline
Dataset2/ & 10088                   & 5388282             \\ \hline
Dataset3/ & 15326 	& 627593             \\ \hline
Dataset4/ &10431                   & 2308279             \\ \hline
Dataset5/ & 18366                   & 2785599             \\ \hline
\end{tabular}
\end{table}

\section{Ranking Rule}

The participants to the competition are ranked based on a dataset and a loss function.

\subsection{Ranking Datasets}

The final ranking was computed based on Dataset5, which was not made public before the ranking. The hidden dataset was different from the provided ones, to avoid overfitting. In order to ensure integrity, the encrypted version of the hidden dataset was uploaded beforehand.

\subsection{Loss Function}
The average error is a loss function, output by \hl{evaluate.py}, it measures how well a program performs on predicting the lines to be replaced. The lower the average line is, the better are the predictions. \\

\noindent
The loss function for one prediction task is $tanh(|\mbox{correct line}-\mbox{predicted line}|)$. The average line error is the loss function over all tasks, as calculated as the average of all individual loss. 
$$\frac{\sum_{i \in Tasks} tanh(|\mbox{correct line}_i-predict(i)|)}{|Tasks|}$$

\noindent
This loss function is designed with the following properties in mind:
\begin{itemize}
\item There is 0 loss when the prediction is perfect
\item There is a bounded and constant loss even when the prediction is far away
\item Before the bound, the loss is logarithmic
\item A perfect prediction is better, but only a small penalty is given to almost-perfect ones. (in our context, some code line replacement are indeed insensitive to the exact insertion locations)
\item The loss is symmetric, continuous and differentiable (except at 0)
\item Easy to understand and to compute
\end{itemize}

\bibliographystyle{abbrv}
\bibliography{references.bib}
\clearpage
\appendix

\section{Command-line Interface}

To play in the competition, a program takes as input a folder name, that folder containing input data files (per the format explained above).

\begin{lstlisting}[language=bash]
$ your-predictor Files
\end{lstlisting}

\noindent
The program outputs on the console, for each task, the predicted line number. Warning: by convention, line numbers start from 1 (and not 0). The program does not have to make prediction for all input files.

\begin{lstlisting}
<Path1> <line number>
<Path2> <line number>
<Path3> <line number>
...
\end{lstlisting}

\noindent
E.g.:
\begin{lstlisting}[breaklines]
/Users/foo/bar/CodRep-competition/Datasets/Dataset1/Tasks/1.txt 42
/Users/foo/bar/CodRep-competition/Datasets/Dataset1/Tasks/2.txt 78
/Users/foo/bar/CodRep-competition/Datasets/Dataset1/Tasks/3.txt 30
...
\end{lstlisting}

\section{Performance Evaluation Tool}

You can evaluate the performance of your program by piping the output to \hl{Baseline/evaluate.py}, for example:

\begin{lstlisting}
your-program Files | python evaluate.py
\end{lstlisting}

\noindent
The output of \hl{evaluate.py} will be:
\begin{lstlisting}
Total files: 15463
Average line error: 0.988357635773 (the lower, the better)
Recall@1: 0.00750177843885 (the higher, the better)
\end{lstlisting}

\noindent
For evaluating specific datasets, use [-d] or [-datasets=] options and specify paths to datasets. The default behaviour is evaluating on all datasets. The path must be absolute path and multiple paths should be separated by \hl{:}, for example:

\begin{lstlisting}[breaklines]
your-program Files | python evaluate.py -d /Users/foo/bar/CodRep-competition/Datasets/Dataset1:/Users/foo/bar/CodRep-competition/Datasets/Dataset2
\end{lstlisting}

\noindent
Explanation of the output of \hl{evaluate.py}:
\begin{itemize}
\item \hl{Total files}: Number of prediction tasks in datasets
\item \hl{Average error}: A measurement of the errors of your prediction, as defined in \textbf{Loss function}. This is the only measure used to win the competition
\item \hl{Recall@1}: The percentage of predictions where the correct answer is in your top 1 predictions. As such, \hl{Recall@1} is the percentage of perfect predictions. We give the recall because it is easily understandable, however, it is not suitable for the competition itself, because it does not has the right properties
\end{itemize}

\section{Provided Baseline Systems}
We provide 5 dumb systems for illustrating how to parse the data and having a baseline performance. These are:
\begin{itemize}
\item \hl{guessFirst.py}: Always predict the first line of the file
\item \hl{guessMiddle.py}: Always predict the line in the middle of the file
\item \hl{guessLast.py}: Always predict the last line of the file
\item \hl{randomGuess.py}: Predict a random line in the file
\item \hl{maximumError.py}: Predict the worst case, the farthest line from the correct solution
\end{itemize}

\noindent
Thanks to the design of the loss function, \hl{guessFirst.py}, \hl{guessMiddle.py}, \hl{guessLast.py} and \hl{randomGuess.py} have the same order of magnitude of error, therefore the value of \hl{Average line error} are comparable.

\end{document}